\documentclass[showpacs]{revtex4}
\usepackage{graphicx}
\usepackage{dcolumn}
\usepackage{bm}
\newcommand{\beq}{\begin{equation}}
\newcommand{\eeq}{\end{equation}}
\newcommand{\bea}{\begin{eqnarray}}
\newcommand{\eea}{\end{eqnarray}}
\newcommand{\bec}{\begin{center}}
\newcommand{\enc}{\end{center}}
\newcommand{\bfr}{\begin{flushright}}
\newcommand{\efr}{\end{flushright}}
\newcommand{\alp}{\alpha}
\newcommand{\om}{\omega}
\newcommand{\tom}{\tilde{\omega}}
\newcommand{\tnu}{\tilde{\nu}}

\newcommand{\kap}{\kappa}
\newcommand{\gam}{\gamma}

\newcommand{\rmi}{{\rm i}}

\begin{document}
\title{Two-photon nonlinearity in general cavity QED systems}
\author{Kazuki Koshino}
 \email{ikuzak@aria.mp.es.osaka-u.ac.jp}
\author{Hajime Ishihara}
\affiliation{
CREST, Japan Science and Technology Agency, 
4-1-8 Honcho, Kawaguchi, Saitama 332-0012, Japan\\
Department of Physical Science, Graduate School of Engineering Science,
Osaka University, Toyonaka, Osaka 560-8531, Japan
}
\date{\today}
\begin{abstract}
We have investigated the two-photon nonlinearity
at general cavity QED systems,
which covers both weak and strong coupling regimes
and includes radiative loss from the atom.
The one- and two-photon propagators are obtained in analytic forms.
By surveying both coupling regimes,
we have revealed the conditions on the photonic wavepacket 
for yielding large nonlinearity depending on the cavity Q-value.
We have also discussed the effect of radiative loss on the nonlinearity.
\end{abstract}
\maketitle

\section{Introduction}

The nonlinear optical response appears strongly
when intense light fields are irradiated onto 
the material with large nonlinearities.
By using an optical cavity,
we can magnify the intensity 
of the input field inside the cavity, 
if the input field is resonant to the cavity mode
\cite{QOpt}.
Thus, we can enhance the nonlinear optical response
by putting the nonlinear material inside the cavity.
This is realized experimentally
by using two-level atoms as the nonlinear material
\cite{Kimble1,Kimble2}.
This idea opened the possibility of obtaining large nonlinearity
even by weak input fields.
In particular, 
the nonlinearity appearing in the two-photon state
is attracting much attention
due to its possible application to quantum information devices
\cite{Nie}
and to recent progresses in the photon-pair manipulation technique
\cite{Take,Eda}.

In order to discuss the dynamics of two-photon state theoretically, 
quantum-mechanical treatment of the photon field is indispensable,
because the nonlinearity appears in the wavefunction of the photon,
not in the amplitude \cite{cm:number}.
Such an analysis was pioneered in a case 
where the nonlinear material is a one-dimensional atom,
which is obtained as 
the lossless and weak-coupling limit of the atom-cavity system
\cite{KojHof}.
However, considering that the photon field is more magnified 
for better (higher Q-value) cavities,
the strong-coupling cases also seem promising 
in yielding large nonlinearity \cite{Ajiki}.
The optimum cavity conditions for the two-photon nonlinearity
have not been sufficiently discussed,
which should be studied with the method applicable to any coupling regime.
In this study, we derive the analytic expression
of the two-photon propagator in general atom-cavity system,
where both weak and strong coupling regimes are covered
and the loss from the cavity 
due to radiative decay into non-cavity modes
is taken into account.
Using this propagator, we discuss the two-photon nonlinearity,
and clarify the optimum condition on the photonic wavepacket 
for achieving large nonlinearity depending on the cavity conditions.

The composition of this paper is as follows.
In Sec.~\ref{sec:2},
the theoretical model of the atom-cavity system is introduced,
and the meanings of the parameters are described.
In Sec.~\ref{sec:3},
we define 
the input and output states of photons.
In Sec.~\ref{sec:4},
the measure of the nonlinearity appearing in the output state
is defined.
In Sec.~\ref{sec:5},
we show the form of the input wavefunction
and a scaling law for this atom-cavity system. 
In Sec.~\ref{sec:6},
we numerically evaluate the nonlinearity for the lossless case,
and clarify the optimum condition for inducing large nonlinearity.
In Sec.~\ref{sec:7},
the effect of the loss is discussed.
The analytic expressions for the one- and two-photon propagators 
are shown in Appendix.

\section{System}\label{sec:2}

\begin{figure}[t]
\includegraphics{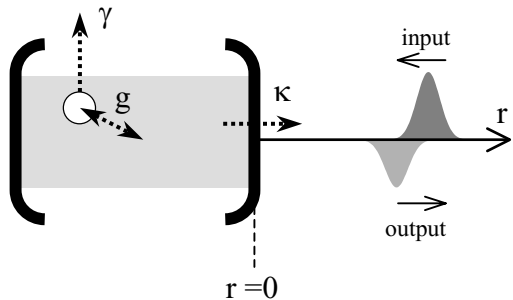}
\caption{\label{fig:sv}
Schematic view of the atom-cavity system.
The right mirror of the cavity is weakly transmissive,
through which the cavity mode is coupled to the external photon field.
$g$, $\gamma$, $\kappa$ represents
the atom-cavity coupling,
the radiative decay rate of the atom into lateral photon modes,
and the decay rate of the cavity mode, respectively.
}\end{figure}

As the nonlinear optical system,
we here investigate a single two-level system 
(hereafter called an ^^ ^^ atom'')
embedded in a one-sided cavity
\cite{CQED}.
The system is schematically illustrated in Fig.~\ref{fig:sv}.
The atom is coupled, besides the cavity mode,
to the photon field in the lateral direction.
The cavity mode is coupled through the right mirror of the cavity
to the external photon field,
which is labeled one-dimensionally by $r$.
Although the external field actually extends only in the $r>0$ region
%
%
and the incoming and outgoing photons are flying in the opposite direction,
we treat the incoming photons to propagate 
in the $r<0$ region in the positive direction 
\cite{Hof2}.

The Hamiltonian of the system is given,
putting $\hbar=c=1$, by
\bea
{\cal H}&=&
\om_a \sigma_+\sigma_-+\int d\mu \ \mu d_{\mu}^{\dagger} d_{\mu}
+\int d\mu \ (\gam_{\mu} \sigma_+d_{\mu}+h.c.)
\nonumber
\\
&+ & \om_c c^{\dagger} c
+\int dk \ k b_{k}^{\dagger} b_{k}
+\int dk \ (\kap_{k} c^{\dagger} b_k+h.c.)
\nonumber
\\
&+ & g(\sigma_+ c + c^{\dagger} \sigma_-),
\eea
where $\sigma_-$, $c$, $b_k$, and $d_{\mu}$ 
are the annihilation operators
for the atom, cavity mode, external photon mode, 
and lateral photon mode, respectively. 
$\om_a$ and $\om_c$ represent the frequencies of 
the atomic transition and the cavity mode,
and $g$ represents the atom-cavity coupling.
Regarding $\gam_{\mu}$ and $\kap_{k}$,
we use the flat-band assumption, i.e.,
$\gam_{\mu}=(\gam/2\pi)^{1/2}$ and $\kap_{k}=(\kap/2\pi)^{1/2}$, 
by which 
the damping rates of the atom and the cavity mode
are given by $\gam$ and $\kap$.

The complex frequencies of the atom and cavity 
are defined by
\bea
\tom_a = \om_a -i\gamma/2, 
\label{eq:toma}
\\
\tom_c = \om_c -i\kappa/2. 
\label{eq:tomc}
\eea
Using these frequencies,
the complex eigenfrequencies 
$\tom_{1,2}$ of the atom-cavity system 
are given by
\beq
(\om-\tom_1)(\om-\tom_2)=(\om-\tom_a)(\om-\tom_c)-g^2.
\label{eq:tom12}
\eeq

We remark here that $\gamma$ is related to the dissipation of this atom-cavity system.
When $\gamma=0$, the inputted photons are always reflected back into the output port,
i.e., the atom-cavity system is lossless.
Contrarily, when $\gamma \neq 0$, 
some of the inputted photons are lost as the spontaneous emission 
into the lateral direction,
i.e., the atom-cavity system is lossy.
We also remark that,
in the limit of $\gamma \to 0$ and $(g,\kappa)\to \infty$ 
keeping $\Gamma=4g^2/\kappa$ constant,
this atom-cavity system is reduced to the one-dimensional atom,
where the atom is directly coupled to the external field 
at $r=0$ with the coupling constant $\Gamma$
\cite{Kimble1,KojHof}.

The real-space operator of the external field, $b_r$,
is given by the Fourier transform of $b_k$;
\beq
b_r=(2\pi)^{-1/2} \int dk e^{\rmi kr}b_k.
\eeq
We note again that the negative (positive) $r$ 
represents the incoming (outgoing) fields.

\section{input and output states}\label{sec:3}

Our main concern in this study is how the initial one- or two-photon wavepackets 
are transformed after the interaction with the atom-cavity system.
In the initial state,
the atom-cavity system is in the ground state
and one or two photons exist in the input port, i.e.,
\bea
|\Psi^{(1)}_{\rm in}\rangle &=& \int dr \psi_{\rm in}(r) b_r^{\dagger}|0\rangle, 
\\
|\Psi^{(2)}_{\rm in}\rangle &=& 2^{-1/2}\int dr_1 dr_2 \psi_{\rm in}(r_1,r_2)
b_{r_1}^{\dagger}b_{r_2}^{\dagger}|0\rangle.
\eea
The one-photon wavefunction $\psi_{\rm in}(r)$ is normalized as 
$\int dr |\psi_{\rm in}(r)|^2 = 1$, and $\psi_{\rm in}(r)=0$ in $r>0$.
Similarly,
$\psi_{\rm in}(r_1,r_2)$ satisfies
$\int dr_1dr_2 |\psi_{\rm in}(r_1,r_2)|^2 = 1$ and
$\psi_{\rm in}(r_1,r_2)=0$ in $r_1,r_2>0$,
and has the symmetry
$\psi_{\rm in}(r_2,r_1) = \psi_{\rm in}(r_1,r_2)$.

Sufficiently after the photons 
have interacted with this atom-cavity system,
the excitations in the atom and the cavity mode
completely 
escape 
to the external modes ($b_k$) 
or to the lateral modes ($d_{\mu}$).
Then, the states of the system are written as
\bea
|\Psi^{(1)}_{\rm out}\rangle &=& \int dr \psi_{\rm out}(r) b_r^{\dagger}|0\rangle 
+ \cdots, 
\\
|\Psi^{(2)}_{\rm out}\rangle &=& 2^{-1/2}\int dr_1 dr_2 \psi_{\rm out}(r_1,r_2)
b_{r_1}^{\dagger}b_{r_2}^{\dagger}|0\rangle 
+ \cdots,
\eea
where the dots imply the terms containing the lateral mode excitations.
While the number of the photons are kept in the output wavepacket
($\int dr |\psi_{\rm out}(r)|^2 = 1$ and 
$\int dr_1dr_2 |\psi_{\rm out}(r_1,r_2)|^2 = 1$)
in case of $\gam=0$,
the output state is attenuated in general cases of $\gam \neq 0$.

The input and the output wavefunctions are related 
by the propagator $G$ as follows;
\bea
\psi_{\rm out}(r) &=& \int dr' G(r;r')\psi_{\rm in}(r'), 
\label{eq:oi1}
\\
\psi_{\rm out}(r_1,r_2) &=& \int dr'_1 dr'_2 
G(r_1,r_2;r'_1,r'_2)\psi_{\rm in}(r'_1,r'_2).
\label{eq:oi2}
\eea
The one- and two-photon propagators are analytically obtainable
for this atom-cavity system,
which are shown in the appendix.
The two-photon propagator has the symmetry of
$G(r_2,r_1;r'_2,r'_1)=G(r_1,r_2;r'_1,r'_2)$,
which guarantees the symmetry in the output wavefunction,
$\psi_{\rm out}(r_2,r_1)=\psi_{\rm out}(r_1,r_2)$.

\section{measure of nonlinearity}\label{sec:4}

When two photons are inputted into this atom-cavity system,
the input wavefunction $\psi_{\rm in}(r_1,r_2)$ 
is finally transformed to the output wavefunction 
$\psi_{\rm out}(r_1,r_2)$.
In order to quantify the nonlinear effect in this process,
we compare $\psi_{\rm out}(r_1,r_2)$ with 
the {\it linear} output wavefunction 
$\psi_{\rm out}^{\rm L}(r_1,r_2)$, which is defined by
\beq
\psi_{\rm out}^{\rm L}(r_1,r_2) = \int dr'_1 dr'_2 
G(r_1;r'_1)G(r_2;r'_2)\psi_{\rm in}(r'_1,r'_2),
\eeq
where $G(r;r')$ is the one-photon propagator.
This linear output is obtained when the atom in the cavity is replaced 
by a harmonic oscillator with the same transition frequency,
i.e., when the nonlinearity of the system is completely removed.

We here define the following quantity $\beta$ by
\beq
\beta=\frac{\int dr_1dr_2 (\psi_{\rm out}^{\rm L})^{\ast}\psi_{\rm out}}
{\sqrt{ \int dr_1dr_2 |\psi_{\rm out}|^2}
\sqrt{ \int dr_1dr_2 |\psi_{\rm out}^{\rm L}|^2 }}.
\label{eq:beta}
\eeq
$\beta$ always lies in the unit circle ($|\beta|\leq 1$) 
due to the Schwartz's inequality,
and $\beta=1$ holds when the response of the system is completely linear
($\psi_{\rm out}=\psi_{\rm out}^{\rm L}$).
Thus, the nonlinear effect is reflected 
in the deviation of $\beta$ from the unity,
and $|\beta-1|$ may be regarded as a measure of the nonlinear effect.
We also note that,
when the atom-cavity system is lossless ($\gamma=0$),
the norms of $\psi_{\rm out}$ and $\psi_{\rm out}^{\rm L}$
in the denominator are always unity, 
and $\beta$ is simply reduced to 
the overlap integral between
$\psi_{\rm out}$ and $\psi_{\rm out}^{\rm L}$.

\section{input wavefunctions and scaling law}\label{sec:5}
Now we embody the above formalism for specific forms of the input states,
and clarify the conditions to obtain large nonlinearity
between two photons.
In this study,
we focus on the case where the two input photons have 
the same wavefunction of the Gaussian form, i.e.,
\bea
\psi_{\rm in}(r_1,r_2)&=&\psi_{\rm in}(r_1-a)\ \psi_{\rm in}(r_2-a),
\label{eq:pin} \\
\psi_{\rm in}(r)&=&(2/\pi d^2)^{1/4}\exp(-r^2/d^2+iqr).
\eea
The input wavefunction is characterized by two parameters,
$q$ (the central frequency) and 
$d$ (the coherent length of the photon). 
$a(<0)$ in Eq.~(\ref{eq:pin}) represents the initial position 
of the photons, which is an irrelevant parameter.

The atom-cavity system is characterized by
$g$ (the coupling between the atom and the cavity mode),
$\om_a-\om_c$ (the frequency mismatch between the atom and the cavity mode),
$\gamma$ (the damping rate of the atom into the lateral modes), and 
$\kappa$ (the damping rate of the cavity mode),
Adding the input-photon parameters $q$ and $d$,
the whole system is specified by the following set of parameters,
$(g, \om_a-\om_c, \gamma, \kappa, q, d)$.
However, the number of the parameters may be decreased
with the help of the scaling law 
that the system with the parameters 
$(\alp g, \alp (\om_a-\om_c), \alp \gamma, \alp \kappa, \alp q, d/\alp)$
is equivalent to the one with 
$(g, \om_a-\om_c, \gamma, \kappa, q, d)$,
where $\alp$ is a positive constant.
Using this law, the system is specified 
by the following set of dimensionless parameters,
$((\om_a-\om_c)/g, \gamma/g, \kappa/g, q/g, gd)$.
$\om_a-\om_c$ is fixed at zero
in the following numerical examples,
and $\om_a$ (=$\om_c$) is chosen as the origin of frequency.

\section{numerical results for lossless cases}\label{sec:6}

In this section, we show the numerical results for $\gamma=0$ cases,
where the atom-cavity system is lossless 
and the number of photons is preserved in the output wavepacket.
The atom-cavity system is then characterized only by the ratio $\kappa/g$,
as has been discussed in Sec.~\ref{sec:5}. 
Eq.~(\ref{eq:tom12}) indicates that the Rabi splitting of the eigenfrequency 
of the atom-cavity system takes place when $\kappa/g < 4$.
Thus, in our definition of the parameters,
the strong (weak) coupling regime is specified by
$\kappa/g \lesssim 4$ ($\kappa/g \gtrsim 4$).

\subsection{Weak coupling regime}
\label{subsec:wcr}

\begin{figure}
\includegraphics{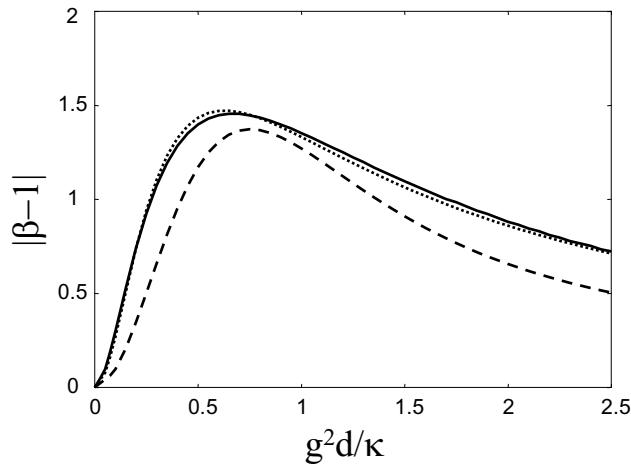}
\caption{\label{fig:wc}
Dependence of $|\beta-1|$ on $g^2d/\kappa$,
where $\kappa/g=10$ (solid line),
$\kappa/g=5$ (dotted line), and
$\kappa/g=2$ (broken line).
The frequency of the photons are resonant to the cavity mode ($q=0$).
}\end{figure}

First, we discuss the weak coupling regime. 
In Fig.~\ref{fig:wc},
the nonlinearity $|\beta-1|$ is plotted 
for the cases of resonant input ($q=0$)
as a function of $\kappa/g$ and $g^2d/\kappa$.
It has been confirmed that the nonlinearity appears most strongly at $q=0$
in the weak coupling regime
and the nonlinearity gets weaker as $|q|$ is increased.

Fig.~\ref{fig:wc} indicates that, roughly speaking, 
the nonlinearity $|\beta-1|$ is dependent solely on $g^2 d/\kappa$
in the weak coupling regime ($\kappa/g \gtrsim 4$).
This is because, in the weak coupling regime,
the atom-cavity system may be regarded as a ^^ ^^ one-dimensional atom'',
where the atom is coupled directly to a one-dimensional photon field
with a coupling constant $4g^2/\kappa$.
It is observed that the nonlinearity is maximized when $g^2 d/\kappa \sim 0.5$.
For example, 
for $g=$120MHz and $\kappa$=900MHz
\cite{Kimble2},
the optimum pulse length $d$ is about 9m.
The qualitative explanation for this optimum condition 
will be given in Sec.~\ref{subsec:op}.

\subsection{Strong coupling regime}
\label{subsec:scr}

\begin{figure}
\includegraphics{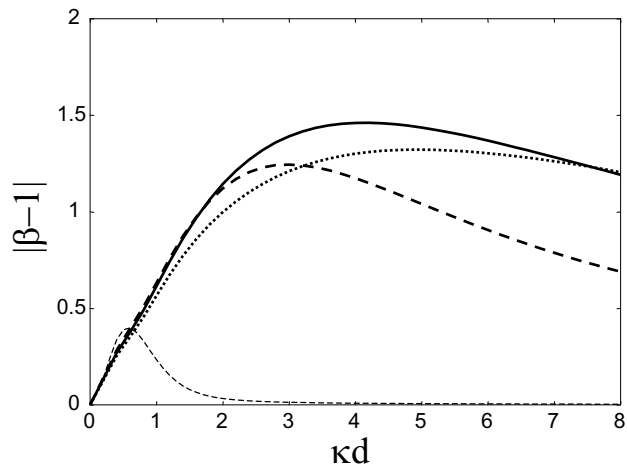}
\caption{\label{fig:sc}
Dependence of $|\beta-1|$ on $\kappa d$, where
$\kappa/g$ is fixed at 0.5 (strong coupling regime).
$q/g=0.8$ (broken line),
$q/g=0.9$ (solid line),
$q/g=1$ (dotted line),
and $q/g=0$ (thin broken line).
}\end{figure}

Next, we discuss the strong coupling regime. 
In Fig.~\ref{fig:sc},
the nonlinearity $|\beta-1|$ is plotted 
for the fixed $\kappa/g$(=0.5)
as a function of $\kappa d$ and $q/g$.
Contrarily to the weak coupling case,
the nonlinearity is weak for the resonant input ($q=0$).
This is because the $q=0$ photons are no more resonant to the cavity mode
due to the Rabi splitting.
Instead, the nonlinearity appears strongly 
when the input photons are tuned to the Rabi-splitted resonant frequency,
$q \sim \pm g$.
The maximum value of $|\beta-1|$ is about 1.5,
which is almost the same as the weak coupling case.
The nonlinearity is optimized for $\kappa d \sim 4$,
the reason of which will be discussed in Sec.~\ref{subsec:op}.

\subsection{Optimum pulse for inducing large nonlinearity}
\label{subsec:op}

In the preceding subsections, we have clarified the optimum frequency $q$ 
and the length $d$ of the photon pulse for inducing large nonlinearity,
in both weak and strong coupling regimes.
Here we explain these optimum conditions from a unified viewpoint.
To this end, we focus on the following wavefunction $\varphi(r)$, 
which is defined by
\beq
e^{i{\cal H}t}\sigma_+|0\rangle =
\int dr \varphi(r) b_r^{\dagger}|0\rangle
\eeq
for large $t(>0)$.
The meaning of $\varphi(r)$ becomes clear 
by multiplying $e^{-i{\cal H}t}$ from the left;
if a single photon pulse $\varphi(r)$ is prepared as an input at time $t_0$,
the photon will completely be absorbed by the atom at time $t_0+t$.
Therefore, it is expected that,
when the input pulse $\psi_{\rm in}(r)$ 
resembles $\varphi(r)$ in shape,
the two input photons try to occupy the atom simultaneously
and the nonlinearity appears strongly.

$\varphi(r)$ has the following form;
\beq
\varphi(r) = 
\left\{
\begin{array}{lr}
\frac{ig\kappa^{1/2}}{\tom_1^{\ast}-\tom_2^{\ast}}
(e^{i\tom_1^{\ast}(r+t)}-e^{i\tom_2^{\ast}(r+t)}) 
& (-t<r<0)
\\
0 & ({\rm otherwise})
\end{array}
\right.,
\label{eq:opt}
\eeq
where $\tom_1$ and $\tom_2$ are 
the complex eigenfrequencies of the atom-cavity system,
defined in Eq.~(\ref{eq:tom12}).
In the weak coupling regime,
$\tom_{1,2}$ are approximately given by 
$-i\kappa/2$ and $-2ig^2/\kappa$.
Noticing that $g^2/\kappa \gg \kappa$ in this regime,
the optimum frequency $q$ and pulse length $d$
are given by
$q \sim 0$ and $d \sim \kappa/2g^2$,
which explains the numerical results in Sec.~\ref{subsec:wcr}.
On the other hand, in the strong coupling regime,
$\tom_{1,2}$ are approximately given by 
$-i\kappa/4 \pm g$.
Therefore, the optimum $q$ and $d$ are
roughly estimated at $q \sim \pm g$ and $d \sim 4/\kappa$,
which are in agreement with the numerical results in Sec.~\ref{subsec:scr}.

\section{numerical results for lossy cases}\label{sec:7}

In the preceding section,
the results for the lossless cases ($\gamma=0$) are presented.
Here, we discuss the lossy cases ($\gamma \neq 0$),
fixing the parameters $\kappa/g=5$ (weak coupling regime)
and $q=0$ for example.

\begin{figure}
\includegraphics{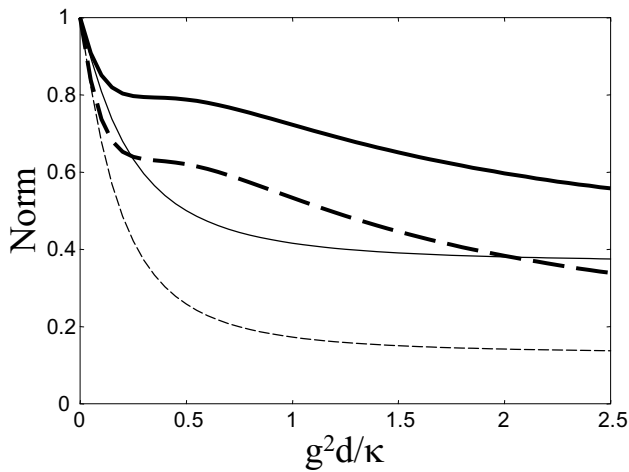}
\caption{\label{fig:norm}
Norms of $\psi_{\rm out}(r_1, r_2)$ (bold lines) and 
$\psi_{\rm out}^{\rm L}(r_1, r_2)$ (thin lines),
for lossy cases of $\gamma/g=0.1$ (solid lines)
and $\gamma/g=0.1$ (broken lines). 
The atom-cavity system is in the weak coupling regime ($\kappa/g=5$)
and the resonant photons ($q=0$) are used.
}\end{figure}

The apparent result of the presence of the loss 
is attenuation of the photon pulses.
In Fig.~\ref{fig:norm}, we have plotted the norm 
of the two-photon output wavefunction
$\psi_{\rm out}(r_1, r_2)$, i.e.,
the probability to find two photons in the output.
The norm of the linear output wavefunction 
$\psi_{\rm out}^{\rm L}(r_1, r_2)$ is also plotted in the same figure.
As expected, the photon pulse is attenuated more significantly
when the loss parameter $\gamma$ is larger,
for both $\psi_{\rm out}$ and $\psi_{\rm out}^{\rm L}$
(compare solid and broken lines).
It is also observed that $\psi_{\rm out}^{\rm L}$
is more attenuated than $\psi_{\rm out}$.
This is understood by the facts that,
in the linear case, the photons are more likely to be absorbed 
by the atom due to the absence of the saturation effect,
and that the loss of photons may take place 
only while the photons are occupying the atom,
as schematically shown in Fig.~\ref{fig:sv}.

\begin{figure}
\includegraphics{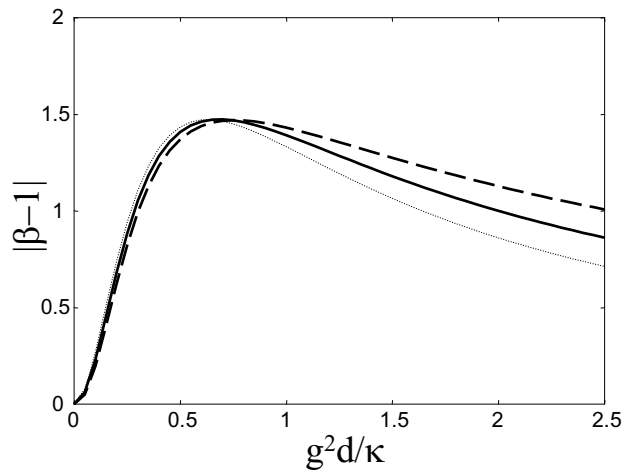}
\caption{\label{fig:lossy}
Effect of the loss on the nonlinearity $|\beta-1|$;
$\gamma/g=0$ (thin dotted line),
$\gamma/g=0.1$ (solid line), and 
$\gamma/g=0.2$ (broken line).
Other parameters are $\kappa/g=5$ and $q=0$.
}\end{figure}

Fig.~\ref{fig:lossy} plots the nonlinear parameter $|\beta-1|$
defined in Eq.~(\ref{eq:beta}),
when the system has finite loss $\gamma$.
It is observed that 
$|\beta-1|$ is slightly decreased (increased) for 
$g^2 d/\kappa \lesssim 0.5$ ($g^2 d/\kappa \gtrsim 0.5$) region but, 
overall, that no qualitative changes are introduced by the presence of the loss.
Thus, although the probability to find two photons in the output 
is largely decreased as shown in Fig.~\ref{fig:norm},
the nonlinear characteristics of the output state 
are almost unchanged from the lossless case
if the output pulse contains two photons.

\section{summary}

We have theoretically investigated a situation 
where two photons are simultaneously inputted into a nonlinear optical system,
and examined the nonlinear effect in the output wavepacket.
As a model nonlinear optical system,
we employed a two-level system (atom)
embedded in a cavity, which is illustrated in Fig.~\ref{fig:sv}.
The one- and two-photon propagators
are obtained in analytic forms,
which are presented in Appendix.
The overlap integral of the linear and nonlinear two-photon output wavepackets,
Eq.~(\ref{eq:beta}), is used to measure the nonlinear effect.
This quantity is numerically evaluated 
both in the weak coupling regime (Fig.~\ref{fig:wc})
and in the strong coupling regime (Fig.~\ref{fig:sc}),
and the conditions for inducing large nonlinearity is revealed.
These conditions are explained 
by the optimum pulse shape, Eq.~(\ref{eq:opt}).
This finding suggests that the pulse shape control 
is more essential in optimizing the two-photon nonlinearity 
rather than the Q-value control of the cavity system.
We also considered the $\gamma \neq 0$ cases,
where the atom-cavity system is lossy.
As shown in Fig.~\ref{fig:norm},
the probability to find two photons in the output 
is largely decreased by the effect of the loss, 
but the nonlinear characteristics of the output two-photon output state
are almost unchanged from the lossless case.

\section*{Acknowledgment}

The authors are grateful for H. Ajiki, M. Bamba, 
and K. Edamatsu for fruitful discussions.

\appendix
\section{propagator}

In this section, we present the analytic forms of 
the one- and two-photon propagators $G(r;r')$ and $G(r_1,r_2;r'_1,r'_2)$.
Using the completeness relation in the one-photon space,
$\hat{1}=\int dr b_r^{\dagger} |0\rangle \langle 0| b_r$,
the one-photon propagator is given by
\beq
G(r;r')=\langle 0|b_r e^{-\rmi {\cal H}t} b_{r'}^{\dagger}|0\rangle,
\eeq
where $r>0>r'$, and 
$t$ is the time difference between the output and the input.
This propagators are derivable by 
standard application of the Green function method
\cite{Mahan}.
Throughout this section, we use the coordinate system moving at the light speed;
transformation to the static coordinate system is done by
replacing $r$ (coordinates without primes) by $r-t$,
or $r'$ (coordinates with primes) by $r'+t$.
In this coordinate system, 
the one-photon propagator is identified as
\begin{widetext}
\bea
G(r, r')&=& G(r-r') = G_0(r-r') + G_2(r-r'), 
\\
G_0(r-r') &=& 
\delta(r-r')-\kap\theta(r'-r)e^{\rmi\tom_c(r-r')}, 
\\
G_2(r-r') &=& 
\kap\theta(r'-r)
\left(
e^{\rmi\tom_c(r-r')}-
\frac{\tom_2-\tom_c}{\tom_2-\tom_1}e^{\rmi\tom_1(r-r')}-
\frac{\tom_1-\tom_c}{\tom_1-\tom_2}e^{\rmi\tom_2(r-r')}
\right),
\eea
\end{widetext}
where 
$\tom_a$ (complex frequency of the atom),
$\tom_c$ (complex frequency of the cavity),
and $\tom_{1,2}$ (complex eigenfrequencies of the atom-cavity system)
are defined in Eqs.~(\ref{eq:toma}), (\ref{eq:tomc}), and (\ref{eq:tom12}).

Next, we proceed to the two-photon case.
Using the completeness relation in the two-photon space,
$\hat{1}=2^{-1}\int dr_1 dr_2 b_{r_1}^{\dagger}b_{r_2}^{\dagger}
|0\rangle \langle 0| b_{r_1}b_{r_2}$,
the two-photon propagator is identified as
$2^{-1}\langle 0|b_{r_1}b_{r_2} e^{-\rmi {\cal H}t} 
b_{r'_1}^{\dagger}b_{r'_2}^{\dagger}|0\rangle$.
This quantity is composed of two kinds of terms;
in the first (second) kind,
the photons initially located at $r'_1$ and $r'_2$
are scattered to $r_1$ and $r_2$ ($r_2$ and $r_1$),
respectively.
With the help of $\psi_{\rm in}(r'_1,r'_2)=\psi_{\rm in}(r'_2,r'_1)$,
it is shown that both of them yield 
the same output wavefunction $\psi_{\rm out}(r_1,r_2)$
after integration in Eq.~(\ref{eq:oi2}).
We can therefore safely regard only the first kind of terms
as the two-photon propagator.
It is given by
\begin{widetext}
\beq
G(r_1,r_2,r'_1,r'_2) =
G(r_1-r'_1) G(r_2-r'_2)
-G_2(r_1-r'_1) G_2(r_2-r'_2)
+\sum_{j=4,6,8}G_j(r_1,r_2,r'_1,r'_2),
\eeq
where $G_j$ ($j=4,6,8$) are defined by
\bea
G_4(r_1,r_2,r'_1,r'_2) &=& 
\frac{-\rmi g^4 \kap^2}{8 \pi^3}
\left[ 
I_4(r_1-r_2, r'_1-r'_2, r_2-r'_1)+
I_4(r_2-r_1, r'_2-r'_1, r_1-r'_2)
\right],
\\
G_6(r_1,r_2,r'_1,r'_2) &=& 
\frac{-\rmi g^6 \kap^2}{8 \pi^3}
\left[ 
I_6(r_2-r_1, r'_1-r'_2, r_1-r'_1)+
I_6(r_1-r_2, r'_2-r'_1, r_2-r'_2)
\right], 
\\
G_8(r_1,r_2,r'_1,r'_2) &=& 
\frac{-\rmi g^8 \kap^2}{8 \pi^3}
\left[ 
I_8(r_1-r_2, r'_1-r'_2, r_2-r'_1)+
I_8(r_2-r_1, r'_2-r'_1, r_1-r'_2)
\right], 
\\
I_4(x,y,z) &=&
\int dk dq d\om
\frac{e^{\rmi k x+\rmi q y+\rmi \om z}J(\om,q,k)}
{(\om-k-q+\rmi\delta)},
\\
I_6(x,y,z) &=&
\int dk dq d\om
\frac{e^{\rmi k x+\rmi q y+\rmi \om z}J(\om,q,k)
(\om-\tom_c-\tom_1)(\om-\tom_c-\tom_2)}
{(\om-k-\tom_c)(\om-q-\tom_c)(\om-\tnu_0)(\om-\tnu_1)(\om-\tnu_2)},
\\
I_8(x,y,z) &=&
\int dk dq d\om
\frac{e^{\rmi k x+\rmi q y+\rmi \om z}J(\om,q,k)}
{(\om-k-\tom_c)(\om-q-\tom_c)(\om-\tnu_0)(\om-\tnu_1)(\om-\tnu_2)},
\eea
where 
\beq
J(\om,q,k)=\frac{1}{(k-\tom_c)(\om-k-\tom_1)(\om-k-\tom_2)
(q-\tom_c)(\om-q-\tom_1)(\om-q-\tom_2)}
\eeq
\end{widetext}
and $\tnu_{0,1,2}$ are defined by
$
(\om-\tnu_0)(\om-\tnu_1)(\om-\tnu_2)
=(\om-\tom_a-\tom_c)
[(\om-2\tom_c)(\om-\tom_a-\tom_c)-2g^2]
$. 
The triple integrals in the definitions of 
$I_{4,6,8}$ are carried out analytically.
It can be confirmed that
the nonlinear part of the Green function,
$G_{\rm NL}=-G_2(r_1-r'_1) G_2(r_2-r'_2)
+\sum_{j=4,6,8}G_j(r_1,r_2,r'_1,r'_2)$,
is nonzero only when 
${\rm max}(r_1, r_2)<{\rm min}(r'_1, r'_2)$
is satisfied, i.e.,
the nonlinearity appears only after the two photons 
have arrived at the atom. 
The one- and two-photon propagators reduce 
to those for the one-dimensional atom
in the limit of $\gamma \to 0$ and 
$(g, \kappa)\to \infty$ keeping $\Gamma=4g^2/\kappa$ constant
\cite{KojHof}.



\begin{thebibliography}{99}

\bibitem{QOpt}
D. F. Walls and G. J. Milburn,
{\it Quantum Optics}
(Springer, Berlin, 1995).

\bibitem{Kimble1}
Q. A. Turchette, R. J. Thompson, and H. J. Kimble,
Appl. Phys. B {\bf 60}, S1 (1995).

\bibitem{Kimble2}
Q. A. Turchette, C. J. Wood, W. Lange, H. Mabuchi, and H. J. Kimble,
Phys. Rev. Lett. {\bf 75} 4710 (1995).

\bibitem{Nie}
M. A. Nielsen and I. L. Chuang,
{\it Quantum Computation and Quantum Information}
(Cambridge Univ. Press, Cambridge, 2000).

\bibitem{Take}
S. Takeuchi, 
Opt. Lett. {\bf 26} 843 (2001). 

\bibitem{Eda}
K. Edamatsu, R. Shimizu, and T. Itoh, 
Phys. Rev. Lett. {\bf 89}, 213601 (2002); 
R. Shimizu, K. Edamatsu, and T. Itoh, 
Phys. Rev. A {\bf 67}, 041805 (2003).

\bibitem{cm:number}
The amplitude (expectation value of the field annihilation operator)
is zero for photon number states.

\bibitem{KojHof}
K. Kojima, H. F. Hofmann, S. Takeuchi, and K. Sasaki, 
Phys. Rev. A {\bf 68}, 013803 (2003);
H. F. Hofmann, K. Kojima, S. Takeuchi, and K. Sasaki, 
{\it ibid} {\bf 68}, 043813 (2003). 

\bibitem{Ajiki}
H. Ajiki, W. Yao, and L. J. Sham,
Superlattices and Microstructures, to be published.

\bibitem{CQED}
{\it Cavity Quantum Electrodynamics},
P. R. Berman, Ed.,
(Academic, San Diego, 1994).

\bibitem{Hof2}
H. F. Hofmann, and G. Mahler,
Quantum Semiclassic. Opt. {\bf 7}, 489 (1995). 

\bibitem{Mahan}
G. D. Mahan, 
{\it Many-Particle Physics},
(Plenum, New York, 1990).

%
\end{thebibliography}
\end{document}